# Composition dependence of electronic, magnetic, transport and morphological properties of mixed valence manganite thin films

Surendra Singh[1,*], J. W. Freeland[2], M.R. Fitzsimmons[3], H. Jeen[4,5] and A. Biswas[4]

[1]*Solid State Physics Division, Bhabha Atomic Research Center, Mumbai 400085 India*
[2]*Advanced Photon Source, Argonne National Laboratory, Argonne, Illinois 60439, USA*
[3]*Quantum Condensed Matter Division, Oak Ridge National Laboratory, Oak Ridge, TN, USA*
[4]*Department of Physics, University of Florida, Gainesville, FL 32611, USA*
[5]*Department of Physics, Pusan National University, Busan 609-735, Korea*

*\*surendra@barc.gov.in*

**Abstract:** We present a comparison of the in-plane length scale over which charge and magnetism are correlated in $(La_{0.4}Pr_{0.6})_{1-x}Ca_xMnO_3$ films with x = 0.33 and 0.375, across the metal to insulator transition (MIT) temperature. We combine electrical transport (resistance) measurements, x-ray absorption spectroscopy (XAS), x-ray magnetic circular dichroism (XMCD), and specular/off-specular x-ray resonant magnetic scattering (XRMS) measurements as a function of temperature to elucidate relationships between electronic, magnetic and morphological structure of the thin films. Using off-specular XRMS we obtained the charge-charge and charge-magnetic correlation length of these LPCMO films near the MIT. The charge-magnetic correlation length (~ 12000 Å) for x = 0.33 was much larger (~4 times) than the charge-charge correlation length (~ 3200 Å) at 20 K. Whereas for x = 0.375 the charge-magnetic correlation length (~ 7500 Å) was smaller than the charge-charge correlation length (~ 9000 Å).

**Introduction:**

Colossal magnetoresistance (CMR) [1] perovskite manganites exhibit numerous phase transitions [2-9], like metal–insulator (MI), ferromagnetic–paramagnetic (FM–PM) and structural phase transitions. A rich variety of electronic and magnetic phases often coexist and compete with one another [1-9] in mixed phase manganites that can be used to achieve interesting functionalities. It is well recognized that many mixed-valence manganites phase separate between the FM metallic phase and the charge-ordered (CO) insulating phases. When driven by disorder near first-order transitions, the length scale of phase-separated domains can be much less than a micron [2, 9-12].

Bulk $(La_{1-y}Pr_y)_{1-x}Ca_xMnO_3$, is a system [6, 11-14] exhibiting phase separation in the range of nanometers to microns for x = 0.33 [9] and 0.375 [11]. Recently Moshnyaga et al., [15] predicted that electronic phase separation (EPS) in thin films of a $(La_{1-y}Pr_y)_{1-x}Ca_xMnO_3$ with x = 0.33 and y = 0.40, develops at the nanometer scale, in which adjacent FM nanodomains are antiferromagnetic (AFM) coupled. The EPS at nanometer length scale at the surface of $(La_{0.4}Pr_{0.6})_{1-x}Ca_xMnO_3$, (LPCMO) film with x = 0.33 was observed using conductive atomic force microscope (cAFM) [16]. Further the macroscopic magnetization, cAFM [16] and transport measurements as a function of applied stress [17] on LPCMO film with x = 0.33 suggested that the direction of the magnetic easy axis and the growth of electronic (metallic) domain at low temperature were different.

Recently using polarized neutron reflectivity we observed that magnetic ordering in LPCMO film with x = 0.33 was not caused by the metal-insulator transition; rather magnetic ordering first occurs at higher temperatures [18]. In addition the results also indicated a magnetic percolation threshold of 0.6 (assuming metallic and ferromagnetic phases are from same parts of the LPCMO film), which is consistent with two-dimensional spin lattices. However, direct evidence for coexistence of magnetic and nonmagnetic regions and their length scales has been elusive in LPCMO films. Here we show that the in-plane length scales over which charge and magnetism are correlated in $(La_{0.4}Pr_{0.6})_{1-x}Ca_xMnO_3$ films with x = 0.33 and 0.375 are different. We also observed that the charge-magnetic correlation length increases below the insulator to metal transition temperature.

**Experimental**

Two single crystal $(La_{0.4}Pr_{0.6})_{1-x}Ca_xMnO_3$ (LPCMO) films with x = 0.33 and 0.375, hence forth known as samples S1 and S2, respectively, were epitaxially grown on (110) $NdGaO_3$ (NGO) substrates in the step-flow-growth-mode using pulsed KrF laser (248 nm) deposition (PLD). During growth, the substrate temperature was 780°C, $O_2$ partial pressure was 130 mTorr, laser fluence was 0.5 J/cm$^2$, and the repetition rate of the pulsed laser was 5 Hz [19]. Previously, scanning transmission electron energy-loss spectroscopy (EELS) microscopy [20, 21] found the composition of an identically prepared film (S1) to be $(La_{1-y}Pr_y)_{1-x}Ca_xMnO_3$(y ~ 0.57, x ~ 0.27) averaged over the entire thickness of the sample. The thicknesses of deposited films were obtained using non resonant (Cu K$_\alpha$ radiation, wavelength = 1.54 Å) X-ray reflectometry.

To study the electronic and magnetic properties of the LPCMO samples, two complementary soft x-ray techniques were used at beamline 4-ID-C of the Advanced Photon Source (APS): XMCD and XRMS [22-24]. The XMCD technique, measured through total electron yield (TEY), probes spin-dependent absorption. The photocurrents (scattering intensity) in XMCD (XRMS) measurements for right (RCP, $I^+$) and left (LCP, $I^-$) circular polarizations of the incident beam were measured independently. Both measurements were taken simultaneously by switching the polarization between LCP and RCP, across the Mn $L_{2,3}$ edges at a fixed incident angle of 10°. Measurements were taken over a temperature range of 20–150 K, using in-plane fields of 700 Oe to saturate the magnetic moment of the sample. The sum $(I^+ + I^-)$ (x-ray absorption spectroscopy, XAS) provides information on the electronic environment of the Mn 3d electrons. While magnetic information is contained in the difference $(I^+ - I^-)$ signal, which in absorption and scattering are referred to as the XMCD and the XRMS, respectively [23, 24]. We also measured the Off-specular (diffuse) XRMS from the samples at the Mn resonant energy of 640.5 eV.

**Results:**

**(a) Transport and conductivity measurements:**

The electrical transport (resistance) measurements were taken using the two-probe method [25] in a closed cycle helium cryostat. The temperature was varied between room temperature to 20 K with an accuracy of better than 0.1 K. Fig. 1 shows resistance normalized to the 300 K resistance [$R(T)/R(300\ K)$] from S1 and S2. We cooled and warmed our samples at a rate of 0.4 K/min. We obtained an insulator to metal (while cooling) ($T_{IM}$) and metal to insulator (while warming) ($T_{MI}$) transition temperatures (location of the peaks of d$R$/d$T$) of 50.7 K and 68.7 K, respectively—a difference of $\Delta T \sim 18$ K for S1. Locations of $T_{IM}$ and $T_{MI}$ are shown as dashed lines in Fig. 1. However, the MIT for S2 occurred at higher temperatures ($T_{IM} = 105.6$ K and $T_{MI} = 109.0$ K) with a smaller thermal hysteresis ~ 4 K.

Temperature dependent morphology and conductance measurements on the surface of similarly grown LPCMO film with x = 0.33 (S1) were reported elsewhere [16]. Figs. 2 (a) and (b) depict the topography with a scan area of 2 μm × 2 μm of identically grown samples S1 and S2, showing different morphologies of the two samples. Distributions of conductivity across the surfaces of these films were measured using conductive atomic force microscopy (cAFM) [16]. To compare the conductivity map we have measured the cAFM images of the surfaces of S1 and S2 at fixed temperature difference of ~20 K below the $T_{MI}$, while warming the samples. Figs. 2 (c) and (d) show the cAFM images at 50 K and 85 K with scan areas of 0.4 μm × 0.4 μm and 2.8 μm × 2.8 μm of the surface of S1 and S2, respectively. The cAFM images show the existence of metallic and insulating phases. The maximum size of metallic domains [Fig. 2(c)] at 50 K for S1 was ~ 1700 Å, whereas for S2 it was ~9000 Å at 85 K [Fig. 2(d)]. We observed even larger metallic domains (~ 12000 Å) for S2 at 50 K (not shown here).

**(b) Hard (non-resonant) X-ray reflectivity**

XRR involves measurement of the x-ray radiation reflected from a sample as a function of wave vector transfer $Q$ (i.e., the difference between the outgoing and incoming wave vectors) [inset (i) of Fig. 3]. In case of specular reflectivity (angle of incidence, $\theta_i$ = angle of reflection, $\theta_f$), $Q = Q_z$ [$= \frac{2\pi}{\lambda}\left(sin(\theta_i) + sin(\theta_f)\right)$, where λ is the x-ray/neutron wavelength] and $Q_x$ [$= \frac{2\pi}{\lambda}\left(cos(\theta_i) - cos(\theta_f)\right)$] = 0, thus we obtain the depth dependent chemical profile of the sample [26]. Whereas in the case of off-specular reflectivity ($\theta_i \neq \theta_f$), the reflectivity (as a function of $Q_x$ at fixed $Q_z$) provides information about the correlation of structure across the sample plane [26-30], e.g. roughness. The specular reflectivity is qualitatively related to the Fourier transform of the scattering length density (SLD) depth profile $\rho(z)$ averaged over the whole sample area [26, 27]. For XRR, $\rho(z)/r_0$, where $r_0$ = 2.82 ×10$^{-5}$ Å is the classical electron radius, can be represented as electron density profile [30].

The specular reflectivities were calculated using the dynamical formalism of Parratt [31], and parameters of the model were adjusted to minimize the value of reduced $\chi^2$ –a weighted measure of goodness of fit [32]. To fit specular XRR data, we considered different model structures consisting of layers representing regions with different electron density. The parameters of the model also included layer thickness and surface roughness/interdiffusion. The specular XRR data, normalized to the asymptotic value of the Fresnel reflectivity ($R_\text{F} = \frac{16\pi^2}{Q^4}$) [26], from S1 (closed triangle) and S2 (closed square) are shown in Fig. 3 (a) along with the best fit (continuous lines). Inset (ii) represents the electron density depth profile of these samples which best fitted the specular XRR. We obtained a thickness of 200±10 and 180±10 Å for LPCMO layer in S1 and S2, respectively, from specular XRR [33].

Investigation of interface morphology of thin film, using the distorted-wave Born approximation (DWBA) formalism, was developed by Sinha et al. [29] which has been subsequently extended to multilayer hetrostructures [34-37]. Using a position sensitive detector (PSD) we collected off-specular (diffuse) reflectivity at different angles of incidence in the scattering plane while diffuse intensity was measured along the length of the PSD. Such a measurement is called a detector scan [35]. Figs. 3 (b) and (c) show the off-specular XRR data as a function of $Q_\text{x}$ at two values of $Q_\text{z,}$ (= 0.17 Å$^{-1}$ and 0.23 Å$^{-1}$) from S1 and S2, respectively. We analyzed the off-specular XRR measurements from the samples under the approximation of self-affine fractal surface, where in-plane height-height correlation function $C(x, y)$ is usually assumed [29, 34-37]: $C(x,y) = <\delta z(0) \delta z(x,y)> = \sigma^2 exp\left(-\left[\frac{\sqrt{x^2+y^2}}{\xi}\right]^{2h}\right)$ ; Where σ is the *rms* value of the surface roughness (correlated roughness), *h* is the roughness exponent, known as Hurst parameter and $\xi$ is the in-plane correlation length of the roughness. The exponent 0<*h*<1 determines the fractal dimension (*D* = 3 –*h*) of the interface (i.e., how jagged the interface is; smoother interfaces have larger values of *h*) [29]. The off-specular XRR were simulated using the formalism developed by Holý et al. [34] to obtain the incoherent diffuse scattering cross-section (see Eq. (16) in ref. [34]). The σ, *h* and $\xi$ for each interface are the parameters of the fit to off-specular XRR data while other parameters obtained from the specular XRR (i.e. thickness and electron density) were kept fixed. We fitted off-specular reflectivity as a function of $Q_x$ at two values of $Q_z$, with the same set of parameters (Table 1). Fits to the off-specular XRR are shown as solid line in Figs. 3 (b) and (c).

**(c) X-ray resonant magnetic scattering**

The XAS and XMCD were calculated by averaging and taking the difference of the photocurrent signals from each polarization, $(I^+ + I^-)/2$ and $(I^+ - I^-)$, respectively. The XMCD is proportional to an average near-surface magnetic moment of a few nanometers [23, 24]. While the XRMS relies on measuring the reflected photon beam, the XRMS is sensitive to the magnetization profile of a few tens of nanometers [23, 24].

XAS and XMCD measurements of the Mn $L$ edge in an applied magnetic field of 700 Oe were taken at an angle of incidence of 10° from the plane of the film's surface. Figs. 4 (a) and (b) show the Mn $L_3$ and $L_2$ edge TEY average absorption $(I^+ + I^-)$ spectra from S1 and S2, respectively, at 150 K (dash lines) and 20 K (solid lines). Similar profiles of XAS spectra have been observed at the intermediate temperature of measurements while warming and cooling the samples. Identical shapes of XAS spectra as a function of temperature suggest that the charge state and local electronic environment of Mn atoms at the surface of both the samples remains unchanged with temperature. Figs. 4 (c) and (d) show the XMCD spectra from S1 and S2, respectively, at 150 K (dash lines) and 20 K (solid lines). The maximum of the (negative) XMCD signal at the Mn $L_3$ edge can be seen at ~640.5 eV, which was the energy used for the off-specular XRMS experiments discussed later. Comparing XMCD data at lowest temperature measured (20 K), in the metallic region (Fig. 1) for both the samples, we observed the magnitude of the near-surface magnetization for the S1 is nearly 5% larger than that of the S2. Figs. 4 (e) and (f) show the XRMS spectra from S1 and S2, respectively, at 150 K (dash lines) and 20 K (solid lines). XRMS spectra from two samples at low temperature are quite different suggesting different magnetization depth profiles for the samples.

The total average magnetization in the near-surface region is proportional to the area bounded by the XMCD spectra [23, 24]. We estimated the temperature dependent magnetization of the samples using XMCD and XRMS spectra. Figs. 4 (g) and (h) show the variation of normalized magnetization from XMCD and XRMS spectra from S1 and S2, respectively, while field cooling. The XMCD signal from S1 decreases faster than the XRMS signal from the same sample, with decreasing temperature. Whereas the opposite behavior was observed for S2. Thus the variations of surface (XMCD) and bulk (XRMS) magnetism of Mn with temperature depend on the chemical compositions and/or on the morphologies of the surfaces.

By measuring the specular reflectivity at maximum XMCD (E = 640.5 eV), hysteresis loops for the Mn moment can be obtained. Figs. 5 (a) and (b) depict the normalized element selective hysteresis loops measured at the $L$3 edge of Mn (maximum XMCD, 640.5 eV) from S1 and S2, respectively, at different temperatures while cooling the samples. We also measured these hysteresis curves while warming the samples. The temperature dependence of the coercive field ($H_c$) of S1 and S2 are shown in the Figs. 5 (c) and (d), respectively. We found thermal hysteresis of $H_c$ about 11K and ~ 0 K for S1 and S2, respectively, which is consistent with the thermal hysteresis of resistance (~18 K and 4 K) for S1 and S2.

### (d) XRMS modeling and Off-specular X-ray resonant magnetic reflectivity

The XRMS, ($I^+ − I^-$), is the charge-magnetic interference term in the scattering amplitude and provides an alternative method for measuring the magnetic dichroism from the subsurface region. In the soft x-ray regime, the longer wavelengths require calculation of the specular intensities using the magneto-optic boundary matrix formalism [38, 39]. The charge-magnetic term in the scattering amplitude can be interpreted as interference between specularly reflected x rays from the chemical and magnetic structures.

Magnetization depth profiles were obtained from the energy-dependent XRMS using a magneto-optical matrix formalism developed by Zak *et al* [38] using the classical dielectric tensor. The formalism requires knowledge of the energy dependence of the refractive index, $n^{\pm} = 1-(\delta_n \pm \delta_m) + i(\beta_n \pm \beta_m)$ of the charge contributions, $\delta_n$ and $\beta_n$, and the magnetic contributions, $\delta_m$ and $\beta_m$. The optical ($\delta_n$, $\beta_n$) and magneto-optic ($\delta_m$, $\beta_m$) constants were obtained from XAS and XMCD measurement using the Kramers-Kronig transform [39, 40]. Figs. 6 (a) and (b) show the optical constants ($\delta_n$, $\beta_n$) as a function of energy obtained from XAS spectra of S1 and S2, respectively at 20 K. Similarly Figs. 6 (c) and (d) show the magneto-optic constants ($\delta_m$, $\beta_m$) as a function of energy as obtained from XAS spectra of S1 and S2, respectively at 20 K. We estimated optic constants and magneto-optic constants for S1 and S2 at different temperatures. Furthermore, we used the chemical structure, i.e. the thickness and roughness parameters, obtained from the non resonant XRR measurements, to analyze the XRMS data at different temperatures. Figs. 6 (e) and (f) show the XRMS data (closed circles) at 20 K and corresponding fit (solid line) from S1 and S2, respectively. The magnetization profile near the surface which best fit the XRMS data at 20 K for S1 and S2 are shown in Figs. 6(g) and (h), respectively. We obtained reduced magnetizations at the surfaces of both films − a result consistent with earlier PNR measurements on similar samples [20, 21].

Diffuse magnetic scattering can arise from both the fluctuation of the magnetic domains [41] and spin [42], which will manifest as magnetic roughness. Since the XRMS measurements were carried out at sufficiently high magnetic fields close to saturation the fluctuation of domain can be neglected and thus off-specular XRMS data can be treated within the same DWBA framework as for the laboratory based x-ray diffuse scattering measurements [43, 44]. The mathematical descriptions of off-specular (diffuse) XRMS, specifically in reflectivity geometry have been discussed in detail elsewhere [45-49]. It has been demonstrated theoretically and experimentally that the diffuse difference ($\Gamma^+ - \Gamma^-$) intensity (charge-magnetic scattering) results predominantly from charge-magnetic cross-correlations while the diffuse sum ($\Gamma^+ + \Gamma^-$) intensity (charge scattering) results predominantly from charge-charge correlations [46, 47]. The experimentally measured diffuse charge scattering data at energy of 640.7 eV, from S1 and S2, at different temperatures while cooling is shown in Figs. 7 (a) and (b), respectively. The diffuse charge scattering data at different temperatures remain similar suggesting that the charge-charge correlation length does not vary with temperature. Similarly Figs. 7(c) and (d) show the experimentally measured diffuse charge-magnetic scattering data at energy of 640.7 eV, from S1 and S2, respectively, at different temperatures while cooling the samples. While the diffuse charge reflectivity is temperature-independent, the diffuse charge-magnetic scattering shows strong temperature-dependence.

Using optic and magneto-optic constants at 640.7 eV estimated from the specular XRMS energy dependent data and other parameters (thickness, number density etc.) obtained from the hard XRR measurements, we fitted the diffuse (both charge and charge-magnetic) XRMS data at different temperatures with different morphological parameters ($\sigma$, $h$ and $\xi$) [43, 44]. The form of the correlation function for both the charge and charge-magnetic fits is that for a self-affine fractal surface with a cutoff length [46], i.e., $C_{cc} = \sigma_c^2 exp\left(-\left[\frac{r}{\xi_{cc}}\right]^{2h_{cc}}\right)$ and $C_{cm} = \sigma_c \sigma_m exp\left(-\left[\frac{r}{\xi_{cm}}\right]^{2h_{cm}}\right)$; where $\xi_{cc}$ and $\xi_{cm}$ are charge-charge and charge-magnetic correlation lengths and $h_{cc}$ and $h_{cm}$ are the corresponding Hurst parameters defining the texture of chemical and magnetic roughness [46]. Figs. 8 (a) and (b) show the diffuse charge scattering data from S1 and S2. The solid lines in Figs. 8 (a) and (b) were fit to diffuse charge scattering data assuming similar morphological parameters (Table 1) as obtained from the non-resonant off-specular XRR data analysis (Fig. 3).

Fig. 9 shows the diffuse charge-magnetic scattering data at a few temperatures while cooling the samples S1 (left panel) and S2 (right panel) in a field of ~ 700 Oe. Solid lines in Fig. 9 are fits to the diffuse scattering data at different temperatures. We obtained smaller correlated root mean square roughness ($=\sqrt{\sigma_c \sigma_m}$) ~ 1.5±0.6 Å as compared to $\sigma_c$, for charge-magnetic correlated surface which did not vary with temperature. The other morphological parameters ($h_{cm}$ and $\xi_{cm}$) obtained from these measurements are plotted as a function of temperature in Figs. 10 (a) and (b) for S1 and S2, respectively. It is also evident from the Fig. 10 that the Hurst parameters obtained from diffuse charge-magnetic XRMS data remains invariant with temperature with average values of $h_{cm} \approx 0.25\pm0.05$ and $0.28\pm0.05$ for S1 [Fig. 10(a)] and S2 [Fig. 10(b)], respectively, - a result again similar to that obtained from charge scattering data for S1 ($h_{cc}$ = 0.25±0.03) and S2 ($h_{cc}$ = 0.20±0.03). The morphological parameters at few temperatures obtained from these measurements are shown in Table 2.

**Discussion:**

The morphological parameters (in-plane correlation length and roughness exponent) obtained from non resonant Cu K$_\alpha$ off-specular XRR are purely due to charge scattering with negligible magnetic contribution. We obtained similar morphological parameters from non resonant Cu K$_\alpha$ off-specular XRR and soft x-ray diffuse sum (I$^+$ + I$^-$) intensity (charge scattering) measurements. We obtained an in-plane charge-charge correlation length ($\xi_{cc}$) of ~3200±400 Å and 9000±700 Å for the LPCMO surface of S1 and S2, respectively, which is much larger as compared to that of the buried substrate-film interface (~ 500±100 Å). A comparison of non resonant Cu K$_\alpha$ off-specular XRR data and corresponding fits assuming different $\xi_{cc}$ for S1 and S2 are depicted in Figs. 11 (a) and (b), respectively, suggest different values of $\xi_{cc}$ for the two samples. In addition both LPCMO surfaces showed a fractal surface with a Hurst parameter (roughness exponent) of ~0.22, which is similar to the Hurst parameters obtained from the diffuse charge-magnetic scattering data. *Thus, the distribution of magnetic moments possesses the same fractal dimension (D ~ 2.7) characteristic of the underlying chemical structure.*

It is evident from Fig. 10 that the in-plane length scale, over which charge and magnetism are correlated, for S1 around 100 K -120 K (above MIT), is the same as the in-plane charge-charge correlation (~ 3200±400 Å) length scale in the insulating phase. A rapid increase of the in-plane charge-magnetic correlation length ($\xi_{cm}$) can be observed at temperature below 70 K. We obtained higher $\xi_{cm}$ (~ 3-4 times of $\xi_{cc}$) for S1 at temperatures below 70 K. Larger charge-magnetic correlation length ($\xi_{cm}$) as compared to charge-charge correlation length ($\xi_{cc}$) for a Fe/Gd multilayer using diffuse XRMS data was also observed earlier [46]. In contrast, we obtained smaller values of $\xi_{cm}$ (~ 2000 – 7500 Å) than the $\xi_{cc}$ (~ 9000±700 Å) for S2. However $\xi_{cm}$ increases with decreasing temperature for both the samples.

Further, to know the correlation of magnetic domain length scales and the length scale over which chemical and magnetic roughness are correlated, we compared the results from cAFM measurements at 50 K and diffuse XRMS data at low temperatures (~ 50 K and 20 K) from S1 and S2. cAFM measurements at 50 K indicate different length scales for the two samples. S1 has smaller ferromagnetic (assuming the metallic phase is ferromagnetic) domains (~ 1700 Å) as compared to that of S2 (~ 11000 Å) at 50 K. $\xi_{cc}$ of S1 (~ 3200 Å) and S2 (~9000 Å) shows a similar trend and suggest that the surface roughness (or charge) are correlated to similar length scales. However, the $\xi_{cm}$ at 50 K for S2 (~ 6000 Å) is smaller than that of S1 (~8000 Å). The results suggest that charge and magnetic roughness are correlated to much higher (lower) length scale in S1 (S2) as compared to the metallic domain area observed by cAFM. Similar trends were observed when we compared the metallic phase obtained from cAFM and $\xi_{cm}$ from diffuse XRMS at similar shifted temperature ($T_{MI}$ – 20 K) for S1 (~ 50 K) and S2 (~85 K).

A typical metallic insulator phase map at low temperature as suggested by cAFM can be depicted in Fig. 11 (c) for S1 and S2. Just below the MIT, the metallic phases in S1 are separated by smaller insulator regions as compared to S2, where we have larger metallic phase separated by large insulator regions. The different values of $\xi_{cc}$ for S1 and S2 are understandable as they show different morphologies, which may be due to different defects (stripes) and doping, *x*, (different internal stress associated with doping) etc., for different films. As temperature decreases below the MIT, the ferromagnetic (metallic) phases of S1 and S2 grow differently because of film morphologies. In the case of S1, we have smaller ferromagnetic domains separated by smaller size non magnetic (insulator) phase (Fig. 11 (c)) which may be reflected in higher values of $\xi_{cm}$ as compared to $\xi_{cc}$. Whereas in the case of S2, the length scales of both metallic (ferromagnetic) and insulating phases are larger (Fig. 11 (c)) as compared to that of S1 and hence show $\xi_{cm} < \xi_{cc}$. Different temperature dependent charge-magnetic correlation and related length scales shown by two samples may also influence the percolation of different phases across the MIT's of these systems. Different percolation of conducting phase while heating and cooling cycle across MIT for LPCMO film grown identically as S1 was also observed earlier by cAFM [16]. Therefore different charge-magnetic correlation across MIT for these systems may be one reason for observing different thermal hysteresis behavior of magnetic properties.

**Conclusion:**

We measured the depth dependence of the chemical and magnetic structures as well as the in-plane charge-magnetic correlation length of the $(La_{0.4}Pr_{0.6})_{1-x}Ca_xMnO_3$ (LPCMO) films with x

= 0.33 (S1) and 0.375 (S2) across the metal insulator transition. We observed reduced surface magnetization for both the LPCMO films compared to the film bulk. The thermal hysteresis in resistance measurements (macroscopic) of the films is correlated with the thermal hysteresis of the coercivity as measured by specular XRMS from these manganite films. Using non-resonant (Cu $K_\alpha$) XRR data, we obtained in-plane charge –charge correlation length of 3200 Å and 9000 Å for S1 and S2, respectively, which were further confirmed by charge diffuse XRMS data. The temperature dependent charge diffuse XRMS data also confirmed that the in-plane charge-charge correlation lengths are independent of temperature. Using magnetic diffuse XRMS data we obtained an increase in the in-plane charge-magnetic correlation length ($\geq$ 5000 Å) below the MIT's for both samples S1 and S2. However, the charge-magnetic correlation length and charge-charge correlation length above the metal to insulator transition were the same for S1. In addition the in-plane charge-magnetic correlation length of S2 is always smaller than its in-plane charge-charge correlation length at all temperatures. The variation of length scale over which the charge and magnetic phases are correlated may influence the percolation of different metallic/magnetic phases across metal insulator transitions of these systems and hence produce different thermal hysteresis of the magnetic properties.

**Acknowledgements:**

This work was supported by the Office of Basic Energy Science (BES), U.S. Department of Energy (DOE), BES-DMS funded by the DOE's Office of BES, the National Science Foundation (DMR 1410237) (AB). Use of the Advanced Photon Source at Argonne National Laboratories was supported by the US DOE under contract DE-AC02-06CH11357. This research partially supported by the Laboratory Directed Research and Development Program of Oak Ridge National Laboratory, managed by UT-Battelle, LLC, for the U. S. Department of Energy.

Table 1: Parameters obtained from non-resonant specular and off-specular XRR

| System | Layer | Thickness (Å) | Correlated roughness (Å) | Correlation length, $\xi$, (Å) | Hurst parameter, $h$ |
|---|---|---|---|---|---|
| S1 | LPCMO | 200±10 | 3±1 | 3200±400 | 0.25±0.03 |
|  | NGO Substrate | - | 5±1 | 500±100 | 0.60±0.10 |
| S2 | LPCMO | 180±10 | 4±1 | 9000±700 | 0.20±0.03 |
|  | NGO Substrate | - | 5±1 | 500±100 | 0.60±0.10 |

Table 2: Parameters obtained from cAFM and diffuse XRMS at low temperature.

| System | Temperature | cAFM | diffuse XRMS | | |
|---|---|---|---|---|---|
|  |  | Size of metallic (ferromagnetic) phase (Å) | In-plane Charge correlation length (Å) | In-plane charge-magnetic correlation length (Å) | Hurst Parameter ($h$) |
| S1 | 20 K | - | 3200±400 | 12000±700 | 0.25±0.02 |
|  | 50 K | 1700 | 3200±400 | 8000±450 | 0.24±0.03 |
|  | 85 K | 1000 | 3200±400 | 3500±300 | 0.27±0.03 |
| S2 | 20 K | - | 9000±700 | 7500±500 | 0.27±0.03 |
|  | 50K | 11000 | 9000±700 | 6000±500 | 0.27±0.03 |
|  | 85 K | 8500 | 9000±700 | 4900±350 | 0.28±0.03 |

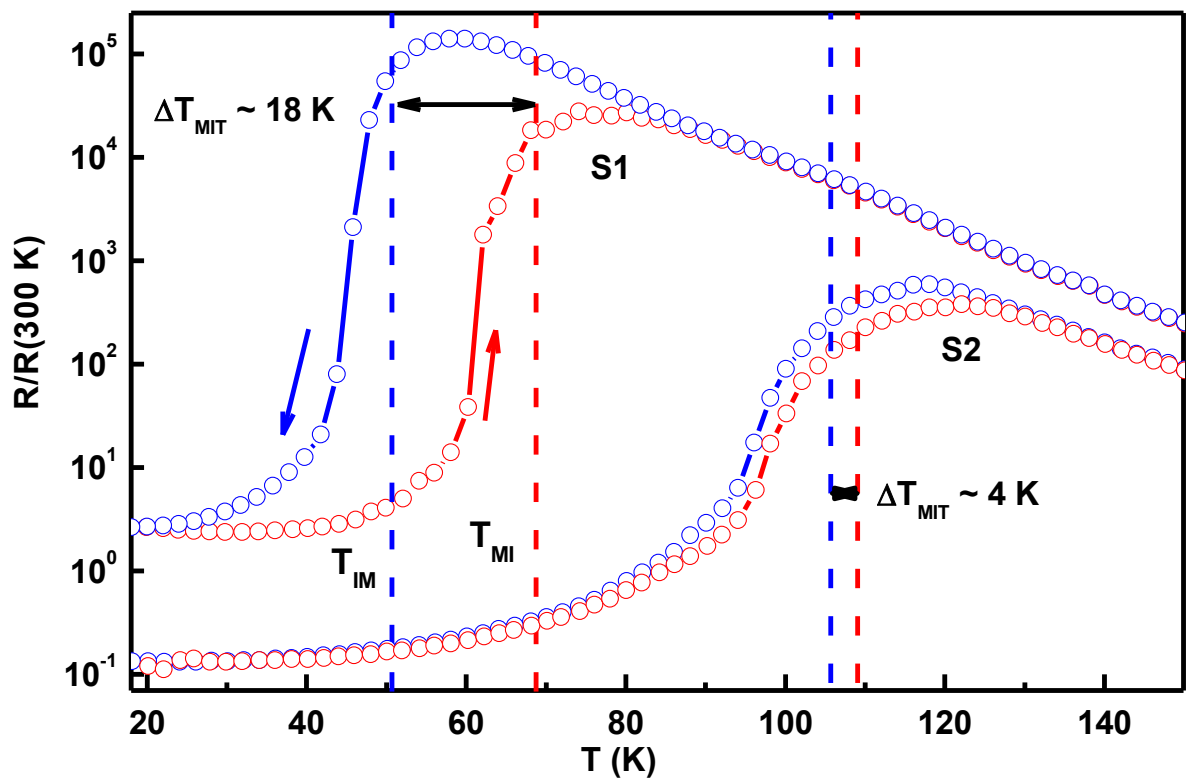

Fig. 1: Electrical transport (resistance, $R$) measurements from samples S1 and S2.

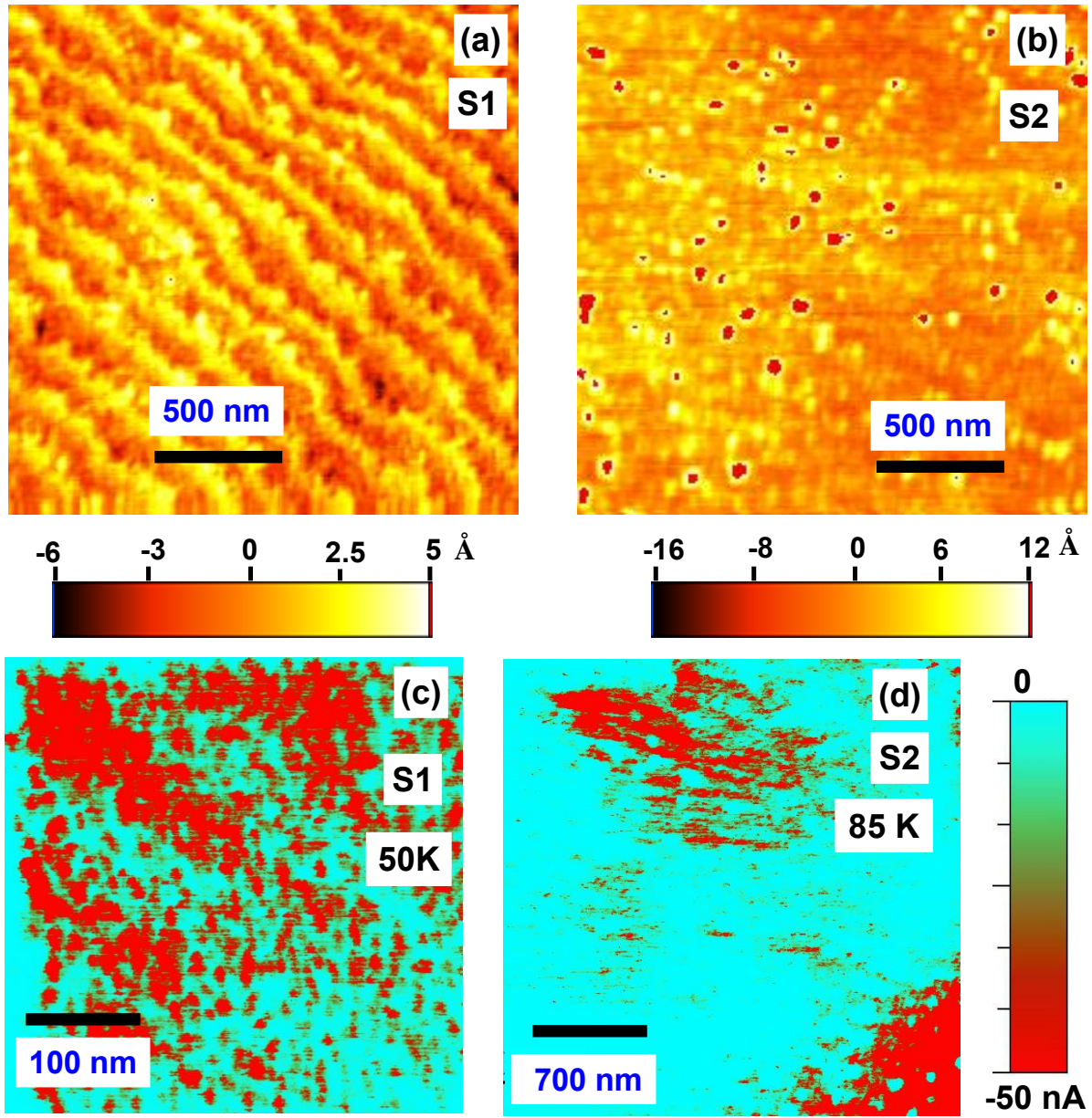

Fig. 2: topography image with a scan size of 2 μm × 2 μm, of the surface of LPCMO samples S1 (a) and S2 (b). (c) and (d) show the current distribution measured by conducting atomic force microscopy for S1 (scan area: 0.4 μm × 0.4 μm) and S2 (scan area: 2.8 μm × 2.8 μm), respectively.

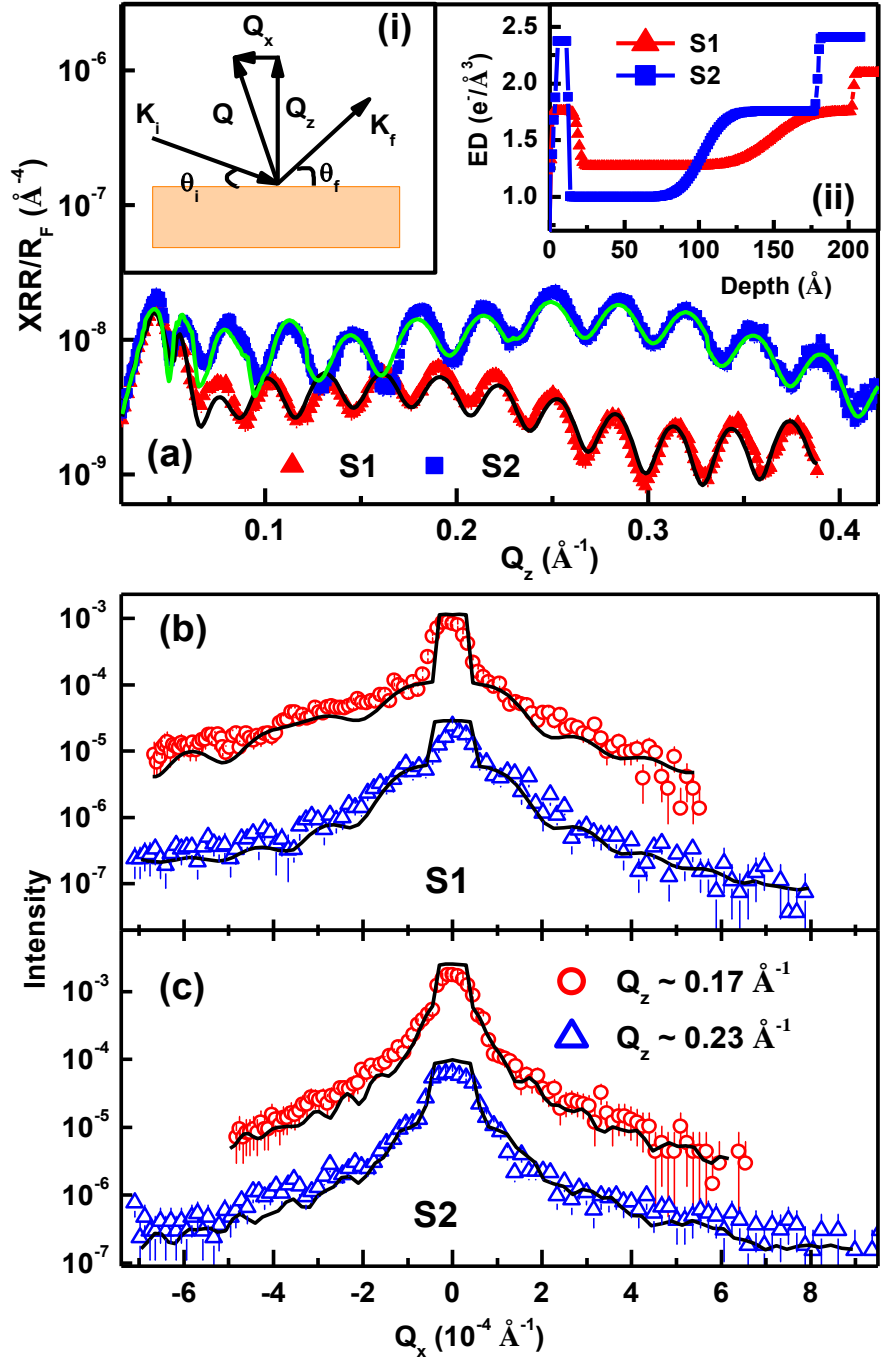

Fig. 3: (a) Specular hard X-ray reflectivity (XRR) data from samples S1 and S2. Inset (i) and (ii) of (a) show the scattering geometry and electron scattering length density (ESLD) depth profiles for S1 and S2, which best fitted (solid lines in (a)) the XRR data. (b) and (c) show the off-specular XRR from S1 and S2, respectively.

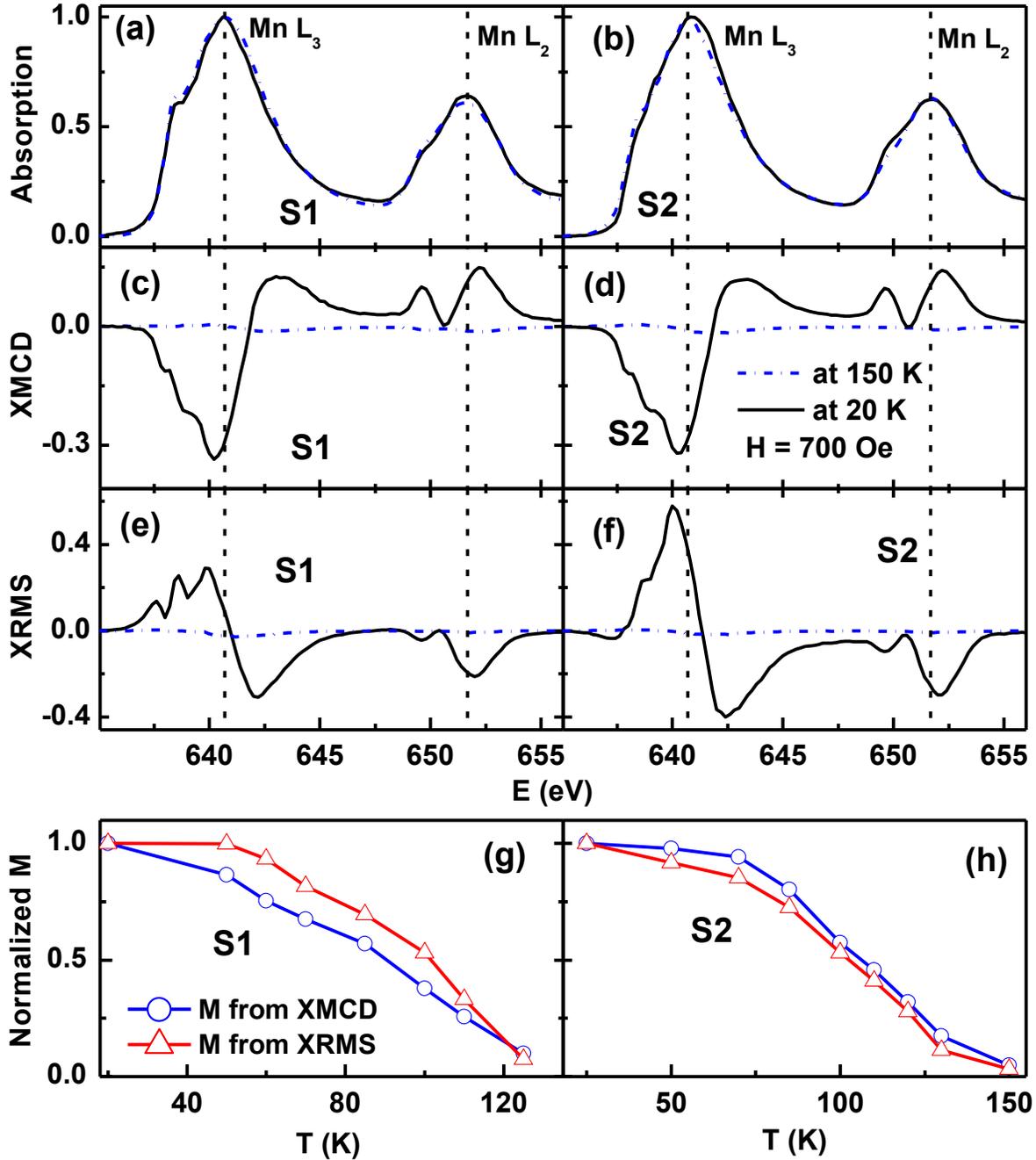

Fig. 4: Near surface x-ray absorption from samples S1 (a) and S2 (b) at 150 K (dash curves) and 20 K (solid curves). XMCD spectra from samples S1 (c) and S2 (d) at 150 K (dash curves) and 20 K (solid curves). XRMS spectra from samples S1 (e) and S2 (f) at 150 K (dash curves) and 20 K (solid curves). (g) and (h) represent temperature dependence of the near surface XMCD peak height (negative) and XRMS peak height for samples S1 and S2, respectively.

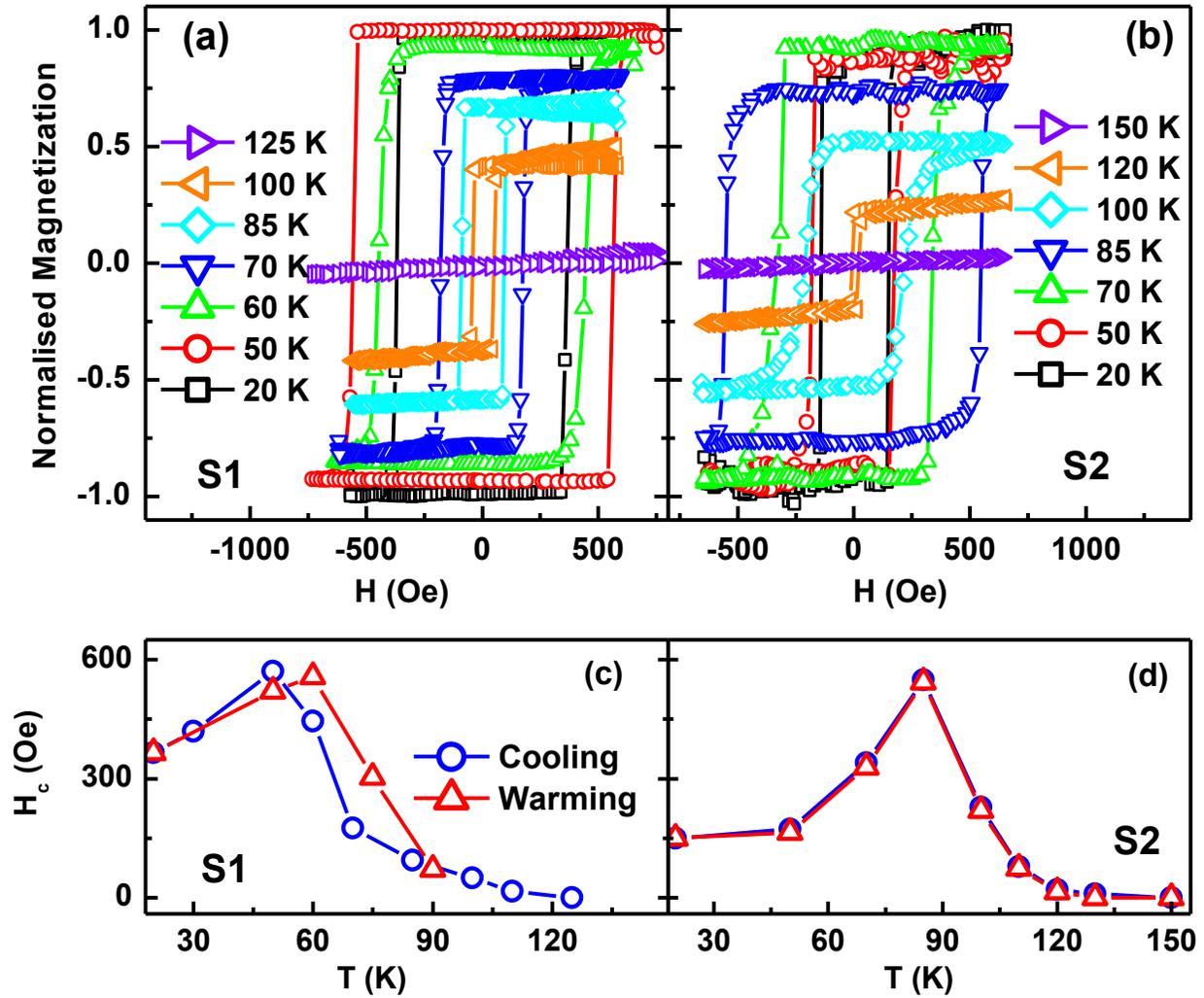

Fig. 5: Normalized element selective hysteresis loops measured at the $L3$ edge of Mn (maximum XMCD, 640.5 eV) from S1 (a) and S2 (2) at different temperatures while cooling the samples. The angle of incidence is 10°. (c) and (d) show the variation of coercive field ($H_c$) as a function of temperature for S1 and S2, respectively.

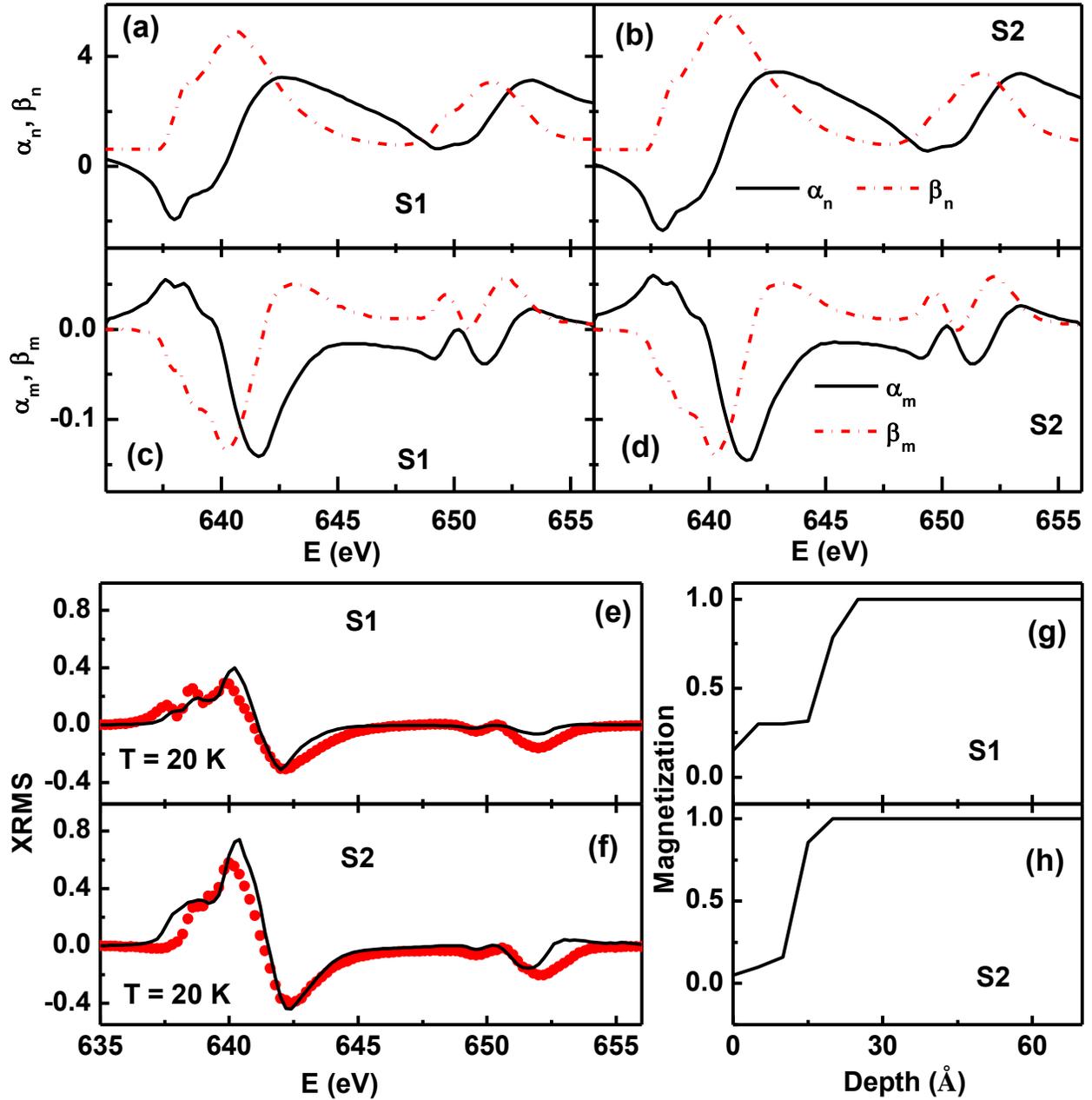

Fig. 6: Optic constants as a function of energy for S1 (a) and S2 (b). Magneto-optic constants as a function of energy for S1 (c) and S2 (d). (e) and (f) show the specular XRMS data (solid circles) at an angle of incidence of $10^0$ and corresponding fits (solid lines) for S1 and S2, respectively, at 20 K. (g) and (h) show the magnetization profile for S1 and S2, respectively which gave best fit to specular XRMS data.

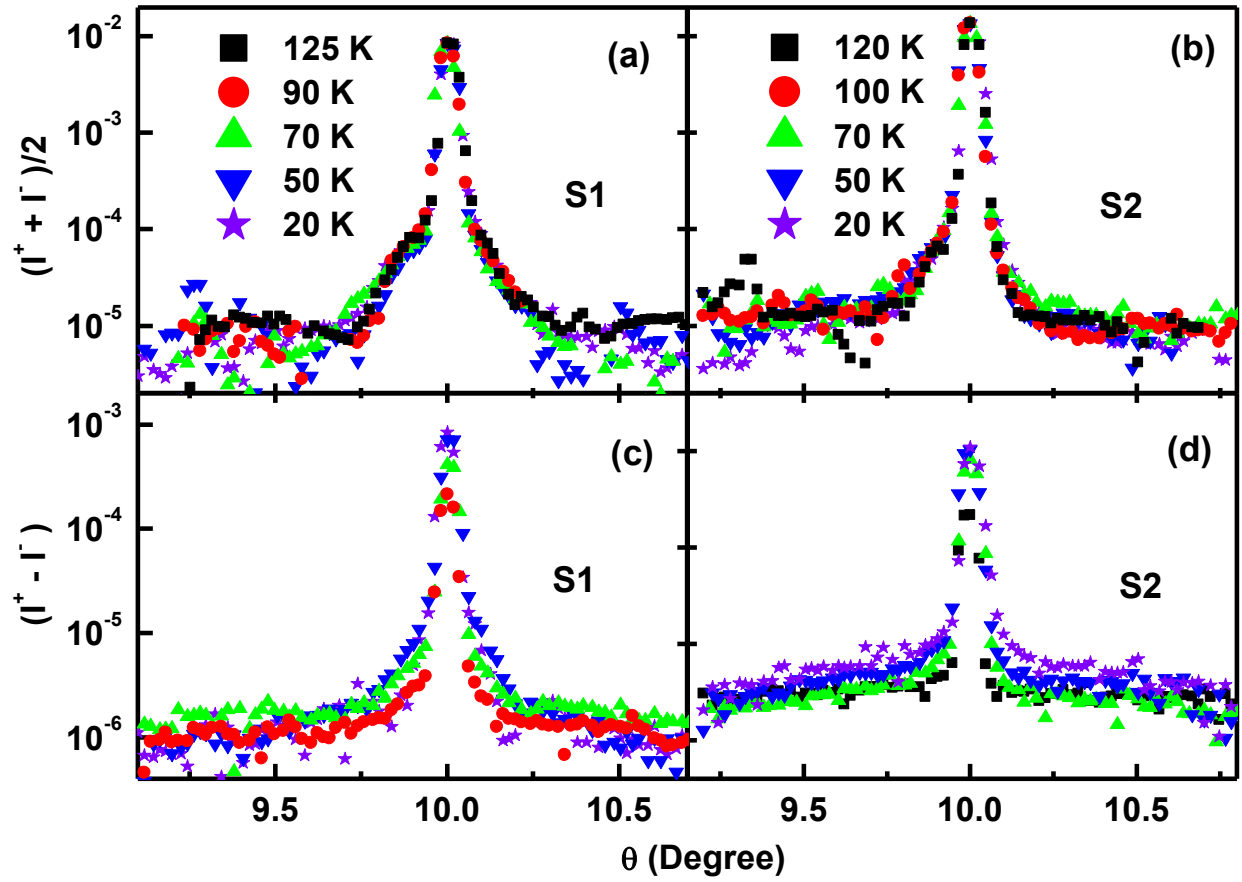

Fig. 7: Diffuse charge scattering data, $(I^+ + I^-)/2$, at different temperatures as a function of angle of reflection from S1 (a) and S2 (b). Diffuse charge-magnetic scattering data, $(I^+ - I^-)$, at different temperatures as a function of angle of reflection from S1 (c) and S2 (d).

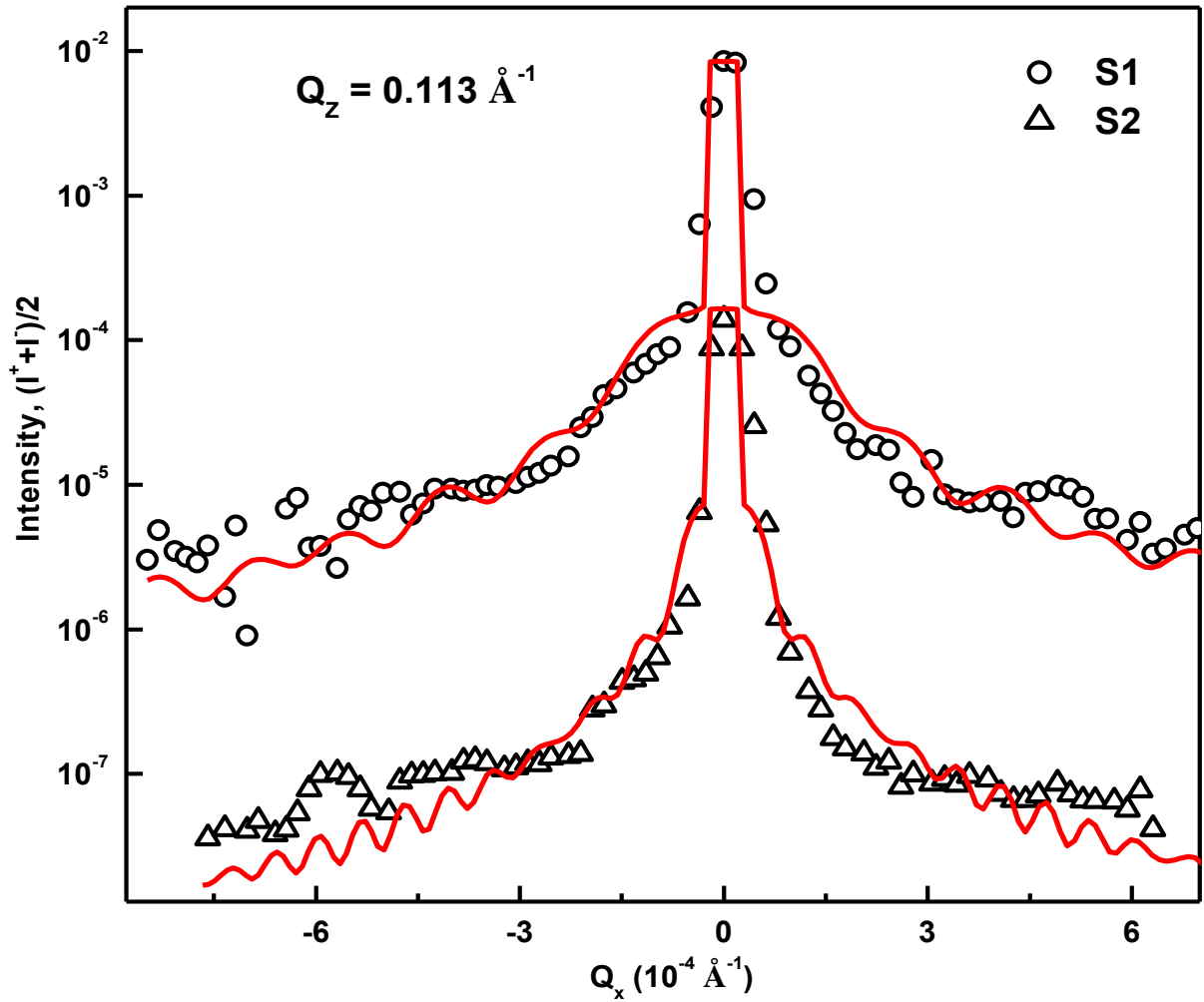

Fig. 8: diffuse charge scattering data, $(I^+ + I^-)/2$, as a function of $Q_x$ from S1 (o) and S2 ($\Delta$). Solid lines are fit to the data. The data have been shifted by 100.

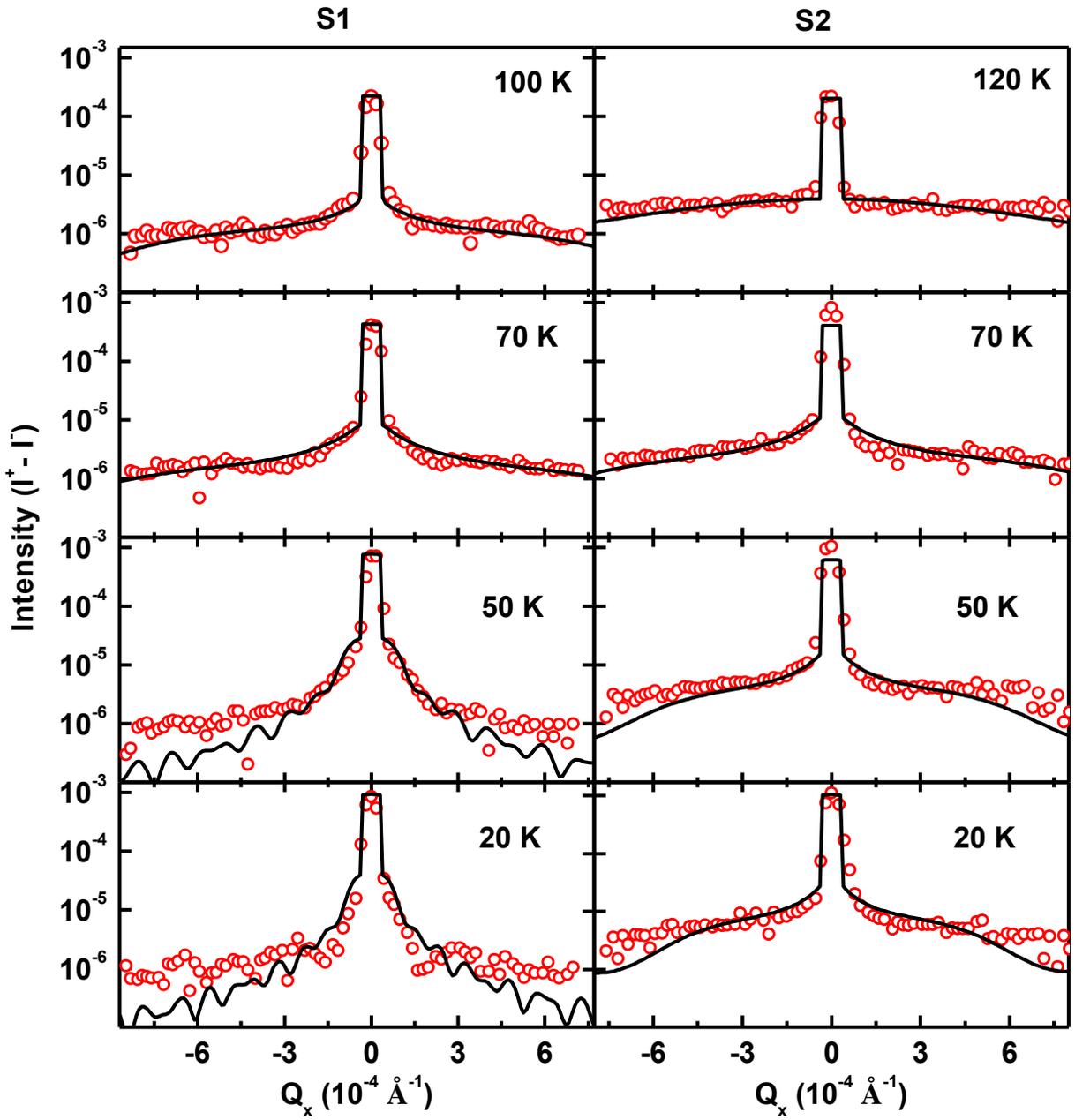

Fig. 9: Diffuse charge-magnetic scattering data (open circles), ($I^+ - I^-$), and corresponding fits (solid lines) at different temperatures as a function of $Q_x$ from S1 (left panel) and S2 (right panel).

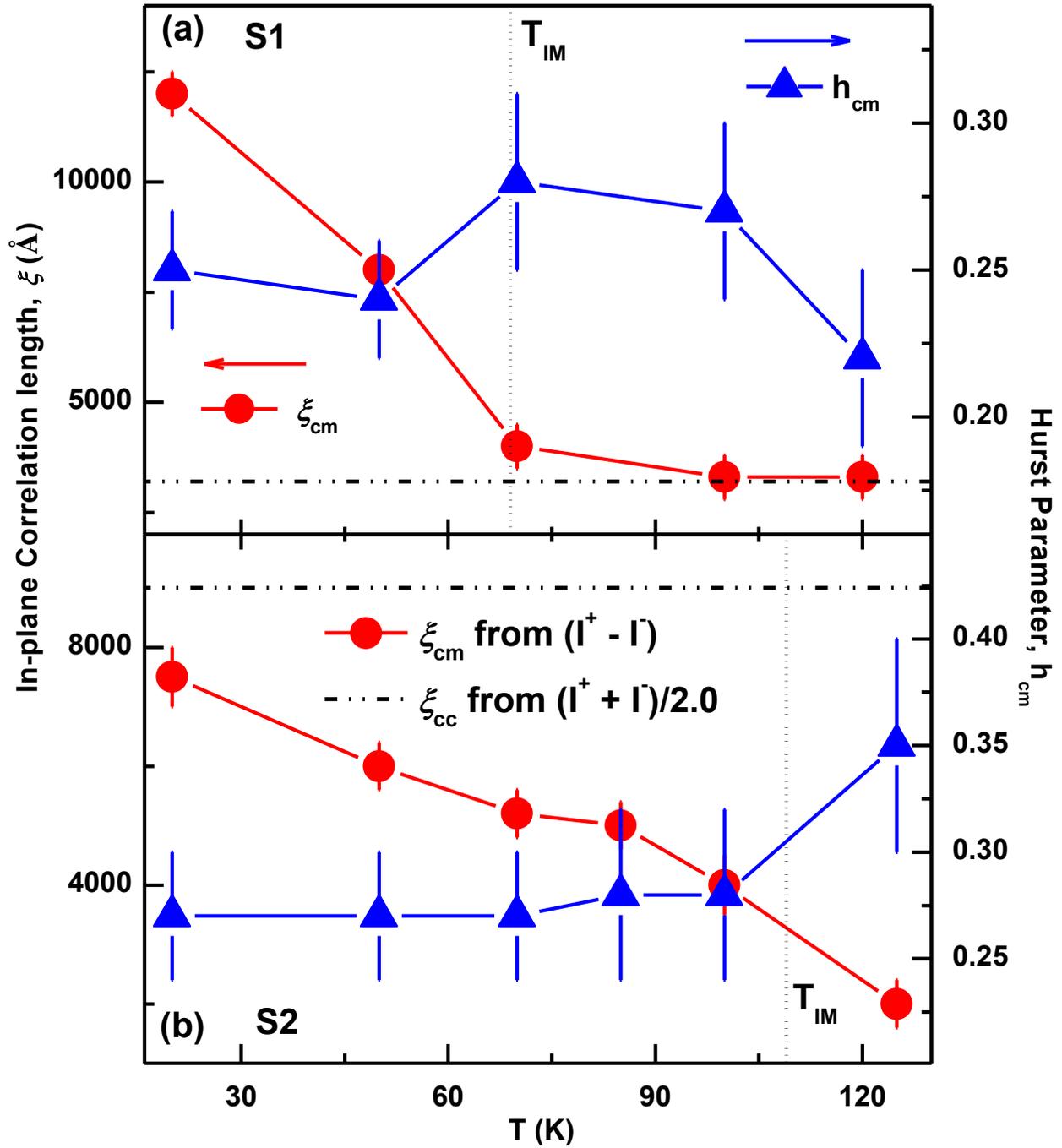

Fig. 10: Morphological parameters (in-plane correlation length, $\xi$, and Hurst parameter, $h$) as a function of temperature obtained from diffuse charge-magnetic scattering data shown in Fig. 9 for S1 (a) and S2 (b).

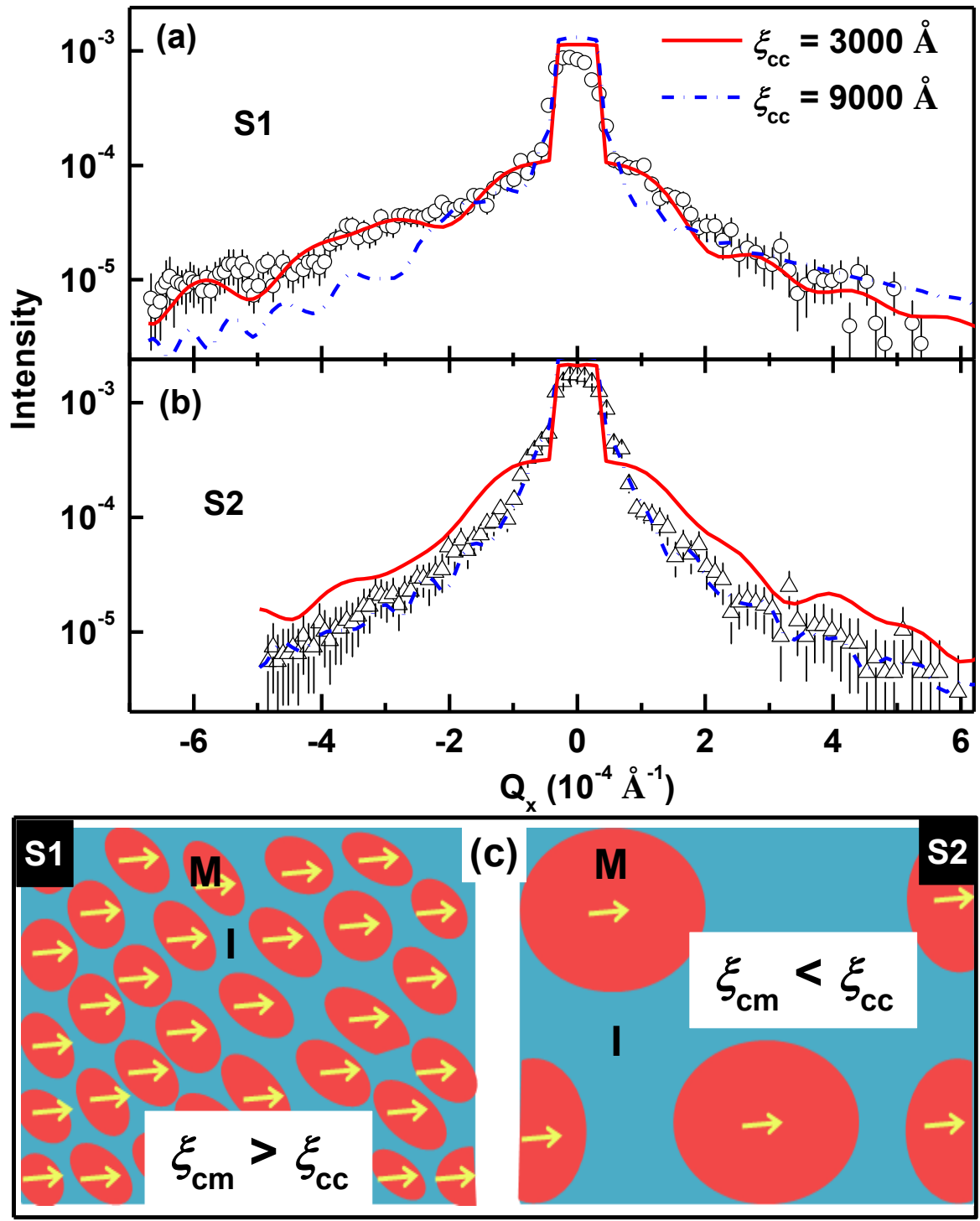

Fig.11: Off-specular XRR (non-resonant) from S1 (a) and S2 (b) and simulated profile with different correlation lengths. (c) shows the schematic of metallic and insulator phase map of S1 and S2 at low temperature.

30